# Highly efficient acceleration and collimation of high-density plasma using laser-induced cavity pressure

J. Badziak [1a)], S. Borodziuk [1], T. Pisarczyk [1], T. Chodukowski [1], E. Krousky [2], K. Masek [2], J. Skala [2], J. Ullschmied [2], Yong-Joo Rhee [3],

[1] *Institute of Plasma Physics and Laser Microfusion, 01-497 Warsaw, Poland*
[2] *PALS Research Centre ASCR, 18200 Prague 8, Czech Republic*
[3] *KAERI, Daejeon, 305 – 353 Korea*

A novel efficient scheme of acceleration and collimation of dense plasma is proposed and examined. In the proposed scheme, a target placed in a cavity at the entrance of a guiding channel is irradiated by a laser beam introduced into the cavity through a hole and accelerated along the channel by the pressure created and accumulated in the cavity by the hot plasma expanding from the target and the cavity walls. Using 1.315-μm, 0.3-ns laser pulse of energy up to 200J and a thin CH target, it was shown that the forward accelerated dense plasma projectile produced from the target can be effectively guided and collimated in the 2-mm cylindrical guiding channel and the energetic efficiency of acceleration in this scheme is an order of magnitude higher than in the case of conventional ablative acceleration.

The ablative pressure of expanding plasma produced at the interaction of laser or X-ray radiation with the outer layer of a fusion target is commonly used to accelerate and compress DT fuel in all currently considered approaches to inertial confinement fusion (ICF) [1, 2]. The ablative acceleration (AA) has also been proposed to be used to accelerate a macroparticle to hypervelocity to ignite the fuel in the impact fast ignition (IFI) fusion scheme [3, 4] as well as for other, non-fusion applications [5]. The energetic efficiency of acceleration, $\eta_{acc}$, defined as the ratio of kinetic energy of the accelerated target to energy of radiation used for the acceleration (producing the ablating plasma in case of AA), is limited by two factors: the absorption coefficient of radiation in the plasma, $\eta_{abs}$, and the hydrodynamic efficiency, $\eta_h$. Even for a short-wavelength radiation (e.g. a 3ω beam of Nd:glass laser), for which these coefficients can be relatively high ($\eta_{abs} \leq 70 – 80\%$, $\eta_h \sim 10 – 20\%$ [1, 2]), the efficiency $\eta_{acc} = \eta_{abs} \eta_h$ is rather low: $\eta_{acc} \leq 7 – 16\%$. Actually, a total acceleration efficiency $\eta_{acc}^t$ – which also takes into account the energy conversion efficiency, $\eta_c$, from the primary laser beam (e.g. a 1ω Nd:glass laser beam) to the short-wavelength radiation (a 3ω beam, X-rays) – is yet lower, and for the Nd:glass laser $\eta_{acc}^t = \eta_{acc} \eta_c \leq 4 – 8\%$, since, usually, $\eta_c \leq 50\%$. One of the important consequences of the low energetic efficiency of acceleration is very high laser energy required for high-gain laser fusion [1, 2, 6].

Fortunately, AA is not the only possible laser method of acceleration of macroparticles or fusion plasmas. About 30 years ago, a cannonball-like acceleration mechanism was proposed to compress a spherical fusion target [7] and its high energetic efficiency was revealed [7, 8]. In its original form, applied in firearms, this mechanism employs the pressure produced in a closed cavity by chemical explosives to accelerate a projectile e.g. in a cannon. In the case of cannonball-like acceleration driven by a laser, the pressure accelerating a projectile (a target) is created in the cavity by the plasma ablating from the laser-irradiated target placed in the cavity [7 – 9]. Such a kind of acceleration can be referred to as laser-induced cavity pressure acceleration (LICPA). Though, for many reasons, the use of LICPA for effective spherical compression of a fusion target is questionable, LICPA has still a great potential to be successfully used for other fusion geometries and/or schemes (e.g. for IFI fusion [3]) as well as for some non-fusion applications.

In this paper, a highly efficient scheme of acceleration and collimation of dense plasma using LICPA is proposed and experimental results demonstrating its very high energetic efficiency are presented.

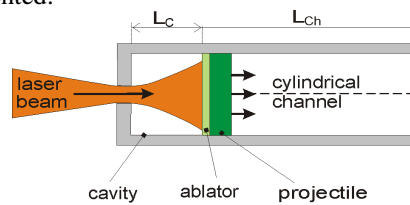

Fig.1. A scheme of laser-driven accelerator using LICPA (see the text).

In the proposed scheme (Fig. 1), a projectile placed in a cavity at the entrance of the guiding channel is irradiated by a laser beam through a small hole in the cavity wall and accelerated along the channel by the pressure produced and accumulated in the cavity by the hot plasma expanding from the irradiated part of the projectile (from the ablator) and from the cavity walls. An important part of the scheme (not considered in previously proposed LICPA schemes) is the guiding

---
[a)] E-mail: badziak@ifpilm.waw.pl

channel, which plays a role similar to that of a barrel in a conventional cannon. In particular, it prevents an "escape" of the pressure from the cavity (which allows for acceleration of the projectile for a long time) and, moreover, it makes it possible to collimate and compress the accelerated plasma.

The motion of a projectile in the cylindrical, one-dimensional LICPA accelerator without losses can be described by the equation:

$$\sigma \frac{d^2z}{dt^2} = p , \quad (1)$$

where p is the pressure in the cavity, $\sigma$ is the areal mass density of the projectile and z is the position of the projectile measured from its initial position. The cavity volume changes during the projectile acceleration as (Fig.1): $V_c = S_c [L_c+(1+\sigma/\Sigma)z]$, where $L_c$ is the initial cavity length, $S_c$ is the projectile (cavity) area, and $\Sigma$ is the areal mass density of the cavity wall. Assuming that the pressure in the cavity changes adiabatically with an increase in the volume, we arrive at the equation:

$$\sigma \frac{d^2z}{dt^2} = \frac{p_0}{[1+(1+\sigma/\Sigma)(z/L_c)]^\gamma} , \quad (2)$$

where $p_0 = (E_L^a / S_c L_c)(\gamma-1)$ is the initial pressure in the cavity, $E_L^a$ is the laser energy absorbed in the cavity and $\gamma$ is the adiabatic exponent. In practical units, the initial pressure can be written as $p_0 \approx 10(\gamma-1)F_L^a / L_c$ , [bar, J/cm$^2$, cm], where $F_L^a = E_L^a / S_c$ is the laser energy fluence in the cavity. From this equation, the following expressions for the projectile velocity and the hydrodynamic efficiency as a function of the distance z of acceleration can be derived:

$$v = v_{max} \left[1-(1+z/z_c)^{1-\gamma}\right]^{1/2} \quad (3)$$

$$\eta_h = \eta_h^{max} \left[1-(1+z/z_c)^{1-\gamma}\right] \quad (4)$$

where:

$$v_{max} = \left[\frac{2F_L^a}{\sigma(1+\sigma/\Sigma)}\right]^{1/2}, \quad \eta_h^{max} = (1+\sigma/\Sigma)^{-1}, \quad (5)$$

and $z_c = L_c(1+\sigma/\Sigma)^{-1}$. It can be seen from (5) that in the case $\sigma << \Sigma$ (the areal mass density of the projectile is much smaller than that of the cavity wall), almost all laser energy absorbed in the cavity can be transformed in the projectile kinetic energy and the hydrodynamic efficiency of acceleration can approach 1, thus it can be significantly higher than in the case of ablative acceleration.

The experiment aimed to demonstrate efficient acceleration and collimation of dense plasma in the proposed scheme was performed at the PALS laser facility [10] in Prague. Both the cavity and the channel of the accelerator were hollowed-out in the massive Al cylinder and their length ($L_c$, $L_{Ch}$) and diameters ($d_c$, $d_{Ch}$) were equal to: $L_c$ = 0.1mm, $L_{Ch}$ = 2mm, $d_c = d_{Ch}$ =  0.3mm (Fig. 2). The diameter of the hole in the cavity wall was $d_{hole}$ = 0.18 mm and the thickness of this wall was 0.1mm. A 10-µm polystyrene (CH) foil placed at the channel entrance was irradiated by a 1.315µm, 0.3-ns laser pulse of energy up to 200 J and intensity up to $10^{15}$W/cm$^2$ injected into the cavity through the hole. The parameters of the forward-accelerated part of the foil (actually – the dense CH plasma), playing the role of a projectile, were measured either in configuration M or N (Fig. 2). The volume of the crater produced in the Al massive target at the output of the channel (configuration M) was assumed to be a measure of energy deposited in the target by forward moving dense plasma and three-frame interferometry [11] as well as ion diagnostics (ion charge collectors) [12] (configuration N) were used to estimate the plasma velocity and temperature. The results of the measurements performed in the LICPA scheme, in particular, the measurements of the crater volume and depth, were compared to the ones obtained for the AA scheme (CH foil accelerated in the channel without the use of the cavity) as well as for a direct laser-Al target interaction.

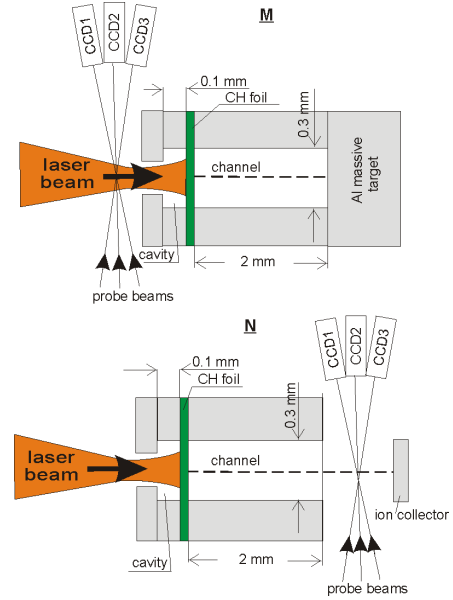

Fig. 2. A simplified scheme of the experimental set-up (out of scale).

The pictures of the craters and the replicas of the craters produced in the Al target in the LICPA scheme, the AA scheme and the L-T scheme (the direct laser-Al target interaction) – obtained at the same laser beam parameters – are shown in Fig. 3. It can be seen that the craters produced with LICPA are considerably larger and deeper than in the case of AA or L-T.

A quantitative comparison of the volume and depth of craters produced in the LICPA, AA and L-T schemes is presented in Figs. 4 and 5. The laser intensity values marked in the figures were calculated assuming that all laser energy (focused to $d_L^h \approx 0.5\, d_{hole}$) was injected into the cavity through the hole in the LICPA scheme. The volume of the craters produced in the Al target by high-density plasma accelerated in the LICPA scheme is more than 30 times greater than that for the AA scheme and 12 – 15 times greater than in the case of direct laser-target interaction. The crater volume for the LICPA scheme increases approximately linearly with an increase in laser energy and the saturation, seen in the plot for the AA scheme, does not appear in the considered energy range. As the crater volume is a measure of the energy deposited in the target and, indirectly, a measure of the kinetic energy of the forward-accelerated plasma, these results demonstrate that the energetic efficiency of acceleration ($\eta_{acc}$) in the LICPA scheme is significantly (an order of magnitude) higher than in the case of ablative acceleration. Also, the energy fluence of the plasma accelerated in the LICPA scheme is remarkably higher than that in the AA scheme, which results in a few times greater crater depth (Fig. 5).

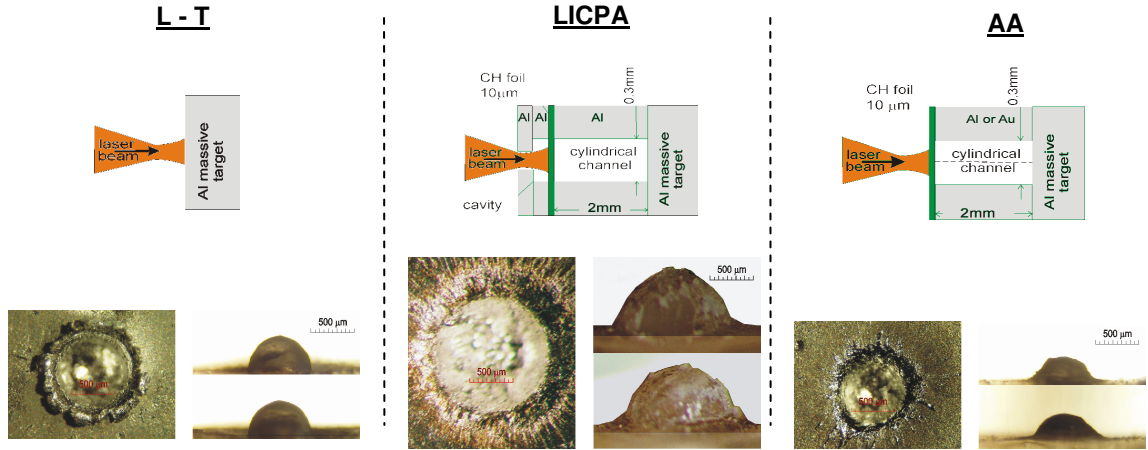

Fig. 3. Craters produced in the Al target by a direct laser-target interaction (L-T) as well as by high-density plasma driven by LICPA or ablative acceleration (AA) and guided in the cylindrical channel. $E_L \approx 130$ J, $\tau_L = 0.3$ ns, $I_L \approx 8 \times 10^{14}$ W/cm$^2$, $L_{Ch}$ = 2mm, $d_{Ch}$ = 0.3mm.

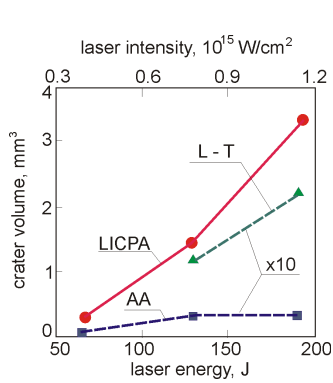

Fig.4. The volume of the craters produced in the Al target by a direct laser-target interaction (L–T) as well as by high-density plasma accelerated in the cylindrical channel by LICPA or AA as a function of laser energy (intensity). Note that the crater volume produced by AA or L – T is magnified 10 times in the figure. $L_{Ch}$ = 2mm, $d_{Ch}$ = 0.3mm.

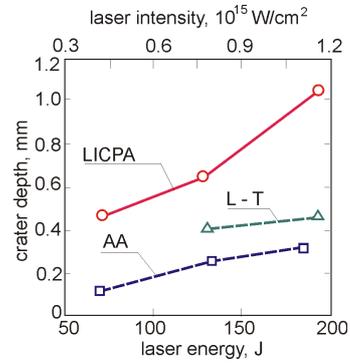

Fig.5. The depth of the craters produced in the Al target by high-density plasma accelerated in the cylindrical channel by LICPA or AA or by a direct laser-target interaction (L – T) as a function o laser energy (intensity). $L_{Ch}$ = 2mm, $d_{Ch}$ = 0.3mm.

It should be underlined that, as it results from our measurements, using a guiding channel in the LICPA scheme is of key importance for production of a fast and dense plasma bunch since it ensures both efficient acceleration and collimation of the plasma. We have observed that in the case of employing LICPA without the channel, a very shallow crater or no crater was produced in the Al target placed at distances from the CH foil comparable to the channel length (1or 2mm). It indicates that in such a case the energy accumulated in the cavity is finally dispersed in a large angle like in the case of using AA for the foil acceleration in free space [13].

The three-frame interferometry using the 2ω PALS laser beam reveals that, at laser energy ~ 130 J, the forward-accelerated plasma covers the distance of 2 mm in ~10 – 15 ns in the channel (Fig. 6), which means that the average plasma velocity in the channel is $<v> \sim (1.5 - 2) \times 10^7$ cm/s. This velocity was found to be comparable to the velocity of a plasma jet of relatively low electron density (~$10^{18}$ – $10^{20}$ cm$^{-3}$) observed at the channel output in the AA scheme [13]. However, contrary to the AA case, such a plasma jet was not recorded in the LICPA scheme. As it results from our numerical simulations using 1D HYDES code, a plausible reason for this difference is the significantly higher density of plasma in the LICPA scheme caused by additional strong plasma compression by the pressure accumulated in the cavity.

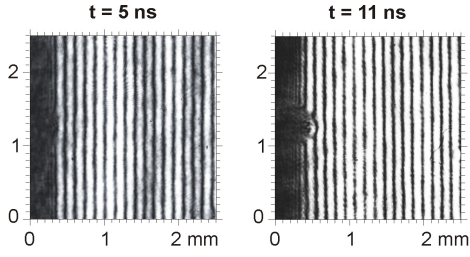

Fig. 6. Interferograms of plasma accelerated in the LICPA accelerator recorded at the output of the cylindrical channel. $E_L \approx 130$ J, $\tau_L = 0.3$ ns, $I_L \approx 8 \times 10^{14}$ W/cm$^2$, $L_{Ch} = 2$mm, $d_{Ch} = 0.3$mm.

The observed high energetic efficiency of acceleration in the LICPA scheme is a consequence of high hydrodynamic efficiency and high laser radiation absorption in the cavity which are higher than in the case of AA or direct L-T interaction. The higher absorption results from the fact that a significant part of laser radiation scattered and reflected back by the ablative plasma can be confined in the cavity and its energy can be transformed in the ablative pressure of the plasma expanding from the irradiated cavity walls. When the area of the hole in the cavity wall is much smaller than the area of the cavity walls, almost all laser radiation can be absorbed in the cavity. In addition, soft X-rays and fast ions emitted from the plasma, which normally are a source of plasma energy loss, are absorbed in the cavity and contribute to the cavity pressure production. For the cavity geometry used in our experiment we can expect that the cavity absorption coefficient is $\eta_{abs} \sim 0.5 - 0.7$. Taking $\eta_{abs} = 0.6$ and moreover: $\sigma_{CH} \approx 1.1 \times 10^{-3}$g/cm$^2$, $\Sigma_{Al} \approx 2.7 \times 10^{-2}$g/cm$^2$, $L_c = 0.1$ mm, $d_c = 0.3$ mm, $E_L = 130$ J, from expressions (5) we obtain for the LICPA scheme: $\eta_h^{max} \approx 0.96$, $v_{max} \approx 4.4 \times 10^7$cm/s and, moreover, $p_o \approx 73$ Mbar. Assuming $z = L_{Ch} = 2$mm and $\gamma = 5/3$, from (3) and (4) we obtain $v \approx 4.2 \times 10^7$cm/s and $\eta_h \approx 0.86$ for the channel output, which result in the acceleration efficiency $\eta_{acc} = \eta_{abs}\eta_h \approx 0.52$ and the average velocity in the channel $<v> \approx 3.6 \times 10^7$cm/s. The last value is a factor 2 higher than the value estimated from the measurement, likely due to simplicity of the model which, in particular, does not take into account any energy losses in the considered LICPA accelerator.

In conclusion, a novel efficient scheme of high-density plasma acceleration and collimation using laser-induced cavity pressure has been proposed and demonstrated. Due to higher hydrodynamic efficiency and higher absorption of laser radiation in the cavity the energetic efficiency of acceleration in this scheme can be an order of magnitude greater than in the case of the conventional ablative acceleration using the "rocket effect". The proposed LICPA accelerator has a potential to be highly useful for fusion-related applications (e.g. for IFI fusion) as well as for other fields of research such as high energy-density physics, laboratory astrophysics or material processing.

This work was supported in part by the HiPER project under Grant Agreement No 211737. The experiment was performed within the Access to Research Infrastructure activity in the Seventh Framework Programme of the EU (Contract No 212025, Laserlab Europe-Continuation)

1. A. Atzeni and J. Meyer-ter-Vehn, Physics of Inertial Fusion, Clarendon Press, Oxford (2004).
2. J. Lindl, Phys. Plasmas **2,** 3933 (1995).
3. M. Murakami and Nagatomo, Nucl. Inst. Meth. Phys. Res. A **544**, 67 (2005).
4. H. Azechi, T. Sakaiya, T. Watari, M. Karasik, H. Saito K. Takeda, H. Hosoda, H. Shiraga, M. Nakai, K. Shigemori, S. Fujioka, M. Murakami, H. Nagatomo, T. Johzaki, J. Gardner, D. G. Colombant, J. W. Bates, A.L. Velikovich, Y. Aglitskiy, J. Weaver, S. Obenschain, S.Eliezer, R. Kodama, T. Norimatsu, H. Fujita, K. Mima, and H. Kan, Phys. Rev. Lett. **102**, 235002 (2009).
5. S. Borodziuk and S. Kostecki, Laser Part. Beams **8**, 241 (1990).
6. E. I. Moses, Nucl. Fusion **49**, 104022 (2009).
7. H. Azechi, N. Miyanaga, S. Sakabe, T. Yamanaka, and Ch. Yamanaka, Jap. J. Appl. Phys. **20**, L477 (1981).
8. K. Yamada, M. Yagi, H. Nishimura, F. Matsuoka, H. Azechi, T. Yamanaka, and Ch. Yamanaka, J. Phys. Soc. Jap. **51**, 280 (1982).
9. S. Borodziuk, A. Kasperczuk, T. Pisarczyk, J. Badziak, T. Chodukowski, J. Ullschmied, E. Krokusy, K. Masek, M. Pfeifer, K. Rohlena, J. Skala, and P. Pisarczyk, Appl. Phys. Lett. **95**, 231501 (2009).
10. K. Jungwirth, A. Cejnarova, L. Juha, B. Kralikova, J. Krasa, E. Krousky, P. Krupickova, L. Laska, K. Masek, A. Prag, O. Renner, K. Rohlena, B. Rus, J. Skala, P. Straka, and J. Ullschmied, Phys. Plasmas **8**, 2495 (2001).
11. A. Kasperczuk, T. Pisarczyk, S. Borodziuk, J. Ullschmied, E. Krousky, K. Masek, K. Rohlena, J. Skala, and H. Hora, Phys Plasmas **13**, 062704 (2006).
12. J. Badziak, J. Makowski, P. Parys, L. Ryć, J. Wołowski, E. Woryna, and A.B. Vankov, J. Phys. D: Appl. Phys. **34**, 1885 (2001).
13. J. Badziak, T. Pisarczyk, T. Chodukowski, A. Kasperczuk, P. Parys, M. Rosiński, J. Wołowski, E. Krousky, J. Krasa, K. Mašek, M. Pfeifer, J. Skala, J. Ullschmied, A. Velyhan, L. J. Dhareshwar, N.K. Gupta, Yong-Joo Rhee, L. Torrisi, and P. Pisarczyk. Phys. Plasmas **16**, 114506 (2009).